\documentstyle[preprint,aps]{revtex}
\tightenlines

\begin{document}
\title{Dynamics of transient disordered vortex states in Bi$_{2}$Sr$_{2}$CaCu%
$_{2}$O$_{8+\delta }$}
\author{B. Kalisky, Y. Bruckental, A. Shaulov, and Y. Yeshurun}
\address{Institute of Superconductivity, Bar-Ilan University, \\
Ramat-Gan 52900, Israel}
\date{July 23, 2003}
\maketitle

\begin{abstract}
The dynamics of transient disordered vortex states in Bi$_{2}$Sr$_{2}$CaCu$%
_{2}$O$_{8+\delta }$ was magneto-optically traced in three experiments: (i)
during continuous injection of transient vortex states while ramping up the
external magnetic field, (ii) during annealing of injected transient states
while keeping the external field constant, and (iii) during annealing of
transient 'supercooled' disordered states while ramping down the external
field. The results reveal front-like propagation (experiment i) or retreat
(experiments ii and iii) of the transient vortex states, at a rate governed
by the rate of change of the external field, the annealing time $\tau $ of
the transient states and the creep rate. The experimental results are
theoretically analyzed in terms of competition between generation and
annealing of transient disordered vortex states. Extraction of the annealing
time $\tau $ from the above three experiments, yields the same results for $%
\tau $ as a function of the induction, $B$, and temperature $T$. Knowledge
of $\tau (B,T)$ \ allows for correct determination of the thermodynamic
order-disorder vortex phase transition line.
\end{abstract}


\section{Introduction}

The nature of the disorder-driven solid-solid vortex phase transition in
high temperature superconductors has been an intriguing issue in physics of
the vortex matter \cite{introduction}. Recent studies have shown that proper
characterization of this transition must take into account effects of
transient disordered vortex states (TDVS) \cite%
{PaltielNature,paltielPRL,GillerPRL,Andrei,gillerJAP,vanderBeekSC,konczy,Kupfer,dimaSC,beenaRapid,beenaBrief,beenaMMM}%
. These transient states are inevitably created by injection of vortices
through inhomogeneous surface barriers while the external magnetic field
increases \cite{PaltielNature,paltielPRL,GillerPRL,beenaRapid,beenaLTup}, or
by ''supercooling'' of the disordered vortex phase while the field decreases %
\cite{vanderBeekSC,dimaSC,beenaRapid,beenaLTdown}. The existence of such
transient states is indicated in time resolved magneto-optical measurements
by the appearance of a sharp change ('break') in the slope of the induction
profiles \cite{GillerPRL,vanderBeekSC}. When the external field is kept
constant, the break moves with time towards the sample edges or center,
indicating annealing of the injected or supercooled TDVS, respectively \cite%
{GillerPRL,beenaRapid,beenaBrief}.

The existence of TDVS near the disorder-driven vortex phase transition,
clarifies several long standing puzzles, such as the apparent increase of
the transition induction with time \cite%
{konczy,Kupfer,beenaRapid,beenaBrief,beenaMMM}, the apparent termination of
the transition line below a certain temperature \cite%
{beenaRapid,beenaBrief,Yesh-Bont,Wen}, yet the appearance of the transition
over a longer time \cite%
{beenaRapid,beenaBrief,beenaMMM,Yesh-Bont,Tamegai,Anders,Wen}, and smearing
of the first order nature of the transition \cite{PaltielNature,paltielPRL}.
Although some aspects of the TDVS have been analyzed \cite%
{dimaTheory,Indenbom}, so far no comprehensive analysis of the dynamics of
TDVS, and its influence on the measured transition, has yet been reported.
At the experimental front, research efforts have been mainly directed
towards eliminating the transient effects, e.g. by utilizing the Corbino
disk configuration in transport measurements \cite{PaltielNature,paltielPRL}
and the vortex dithering technique in magnetic measurements \cite%
{AvrahamNature}. These sophisticated experiments successfully uncovered the
underlying thermodynamic order-disorder vortex phase transition. In this
work we investigate, both experimentally and theoretically, the dynamics of
the transient disordered vortex states, and utilize this knowledge to
extract the thermodynamic order-disorder vortex phase transition line. We
present comprehensive time resolved magneto-optical measurements in Bi$_{2}$%
Sr$_{2}$CaCu$_{2}$O$_{8+\delta }$ (BSCCO), demonstrating continuous
injection of transient vortex states while ramping up the external magnetic
field, annealing of injected transient states while keeping the external
field constant, and annealing of transient 'supercooled' disordered states
while ramping down the external field. In all these experiments the TDVS
exhibit front-like propagation, or retreat, with a velocity depending on the
rate of change of the external field, the creep rate and the annealing time $%
\tau $ of the transient states. The dynamic behavior of the front is
quantitatively analyzed in terms of competition between two mechanisms,
namely generation and annealing of TDVS. This analysis enables extraction of
the annealing time $\tau $, characterizing the TDVS, as a function of the
induction $B$ and temperature $T$. We show that knowledge of $\tau (T,B)$
enables the extraction of the thermodynamic vortex order-disorder transition
line, $B_{od}(T)$. The extracted $B_{od}(T)$ line is significantly different
from the apparent transition lines commonly measured from the onset of the
second magnetization peak, ignoring effects of TDVS.

\section{Experimental}

Measurements were performed on a 1.55$\times $1.25$\times $0.05 mm$^{3}$ Bi$%
_{2}$Sr$_{2}$CaCu$_{2}$O$_{8+\delta }$ single crystal ($T_{c}=$ 92 K). The
crystal was grown using the traveling solvent floating zone method \cite%
{Motohira}. This crystal was specially selected for its uniformity of flux
penetration and was checked by magneto-optical imaging before and after it
was cut into a rectangle. In the course of measurements, magneto optical
(MO) snapshots of the induction distribution across the sample surface were
recorded at time intervals of typically 40 ms, using iron-garnet MO
indicator with in-plane anisotropy \cite{Vlasko-review} and a high speed CCD
camera (Hamamatsu C4880-80).

\subsection{Field Sweep Up (FSU)}

The process of injection of TDVS through the sample edges by sweeping up the
external field, was magneto-optically recorded at different sweep rates. In
these experiments, the sample was cooled down to the measuring temperature
in zero field, then the external magnetic field, $H_{ext}$, was ramped up at
a constant rate between 4 and 1600 G/sec, from zero to about 850 G. While
the external magnetic field was ramped up, snapshots of the induction
distribution across the crystal surface were taken successively at constant
field intervals (usually 10 G).

Figure 1 shows the induction profiles across the crystal width deduced from
the magneto-optical images, taken at $T=$ $23$ $K$, while the external field
was ramped up at a rate of 53 G/sec. When $H_{ext}$ reaches a value of
approximately $430$ $G$, a sharp change in the slope of the profile
(''break'') appears at $B_{f0}$ $\simeq $ $360$ $G$, indicating coexistence
of two distinct vortex states, characterized by high and low persistent
current densities: A high persistent current state near the sample edges and
a low persistent current state near the center \cite{GillerPRL}. The high
persistent current state is identified as a TDVS, because it decays with
time when the external field is kept constant (see Sec. IIB below). When the
external field is continuously increased, the break moves towards the sample
center, indicating propagation of the transient disordered state\ front
deeper into the sample. At the same time, the induction at the break
increases monotonically with a rate decreasing with time. As shown below
(see Sec. IIIA), these results can be explained in terms of competition
between injection and annealing processes of TDVS. While the rate of
injection remains approximately constant, the rate of annealing is high for
low inductions, and decreases sharply as the induction increases towards $%
B_{od}$. The first appearance of the break indicates a stage where the
injection rate starts to overcome the annealing rate.

Increasing the field sweep rate, or decreasing temperature at a constant
sweep rate, shifts $B_{f0}$ downwards, indicating that the injection process
of TDVS starts to overcome the annealing process at lower inductions. The
effect of increasing sweep rate is demonstrated in Figure 2, which shows the
induction profiles measured at $T=$ 25K while ramping up the external field
at 4, 160, and 800 G/sec. A rate increase from 4 to 160 G/sec, shifts $B_{f0}
$ down from $\symbol{126}$450 to $\symbol{126}$400 G. A further increase of
the rate to 800 G/sec shifts $B_{f0}$ down to $\symbol{126}$330 G. The
effect of temperature on $B_{f0}$ is demonstrated in Figure 3. For the same
sweep rate of 160 G/sec, lowering temperature from $25$ $K$ to $23$ $K$,
shifts $B_{f0}$ down from $\symbol{126}400$ G to $\symbol{126}330$ G. As
temperature is further lowered to $21$ $K$, $B_{f0}$ is shifted down to $%
\symbol{126}200$ G. Figure 4 shows the velocity of the TDVS front as a
function of time for different sweep rates, measured at $T=23$ $K$. The
initial velocity of the front (the first appearing break) is zero. It then
accelerates with a decreasing rate, approaching a constant velocity
determined by the rate of change of the external field.

\subsection{\protect\bigskip Annealing at a constant field}

The annealing process of injected TDVS was magneto-optically recorded while
keeping the external magnetic field at a constant level. Injection of TDVS
throughout the sample was accomplished by abruptly raising the external
field to a target value between 140 and 850 G (rise-time 
\mbox{$<$}
50 ms). Immediately after reaching the target value of the external field,
magneto-optical snapshots of the induction distribution across the sample
surface were recorded at time intervals of 40 ms for 4 seconds, and 300 ms
for additional 26 seconds.

Figure 5 shows the time evolution of the induction profiles at T= 21 K after
abruptly increasing the external field from zero to 465 G. Initially, the
profiles are smooth, without a break, indicating a single vortex phase
throughout the sample. However, after approximately 0.5 second, a break
appears in the slope of the induction profiles, progressing with time
towards the sample edge. The break, (a typical one is marked in the figure
by an arrow), separates between a high persistent current state near the
sample edge and a low persistent current state near the center. The high
persistent current region near the edge shrinks with time, and therefore the
vortex state in this region is identified as a transient disordered state.
The vortex state in the expanding, low persistent current region near the
center, is identified as the thermodynamic quasi-ordered phase \cite%
{GillerPRL}. The front of the growing thermodynamic phase moves initially
with a large velocity and decelerates with time to zero velocity as shown in
Figure 6, for $H_{ext}=500$ and $625$ $G$. After reaching zero velocity, the
movement of the front reverses its direction, indicating an end of the
annealing process and beginning of injection of TDVS into the sample
interior. One would expect \cite{GillerPRL} that this turning point is
obtained when the induction at the front reaches the value of the
thermodynamic vortex order-disorder transition induction $B_{od}$. However,
the data shows that the annealing process ceases at an induction of $\symbol{%
126}$ $400$ $G$, i.e. below $B_{od}=430$ $G$ (see Section IV).

We show below (Section IIIB), that the origin of this phenomenon is
associated with the fundamental difficulty of realizing a 'pure' annealing
experiment; While the external field is kept constant, disordered vortex
states are continuously injected through the sample edges via flux creep.
For inductions well below $B_{od}$ the annealing rate is much larger, thus
the thermodynamic quasi-ordered vortex state continuously grows. However, as
the induction at the front approaches $B_{od}$, the annealing rate
continuously decreases until a point is reached when the rate of injection
of TDVS due to flux creep equals the annealing rate. At this point\ in time
the growth of the thermodynamic quasi-ordered phase comes to an halt. After
this point, the rate of injection of TDVS due to flux creep becomes larger
than the annealing rate, and as a result TDVS are injected into the sample.
This is manifested by the reverse movement of the front towards the sample
center.

\subsection{Field Sweep Down (FSD)}

''Supercooling'' of a disordered vortex phase \cite{vanderBeekSC} and
annealing of the transient supercooled state were magneto-optically traced
in field sweep down experiments. In these experiments, an external field of
850 G was initially applied for long enough time to ensure establishment of
a disordered vortex phase. The field was then ramped down to zero at a
constant rate between 4 and 1600 G/sec. While the external field was ramped
down, snapshots of the induction distribution across the crystal surface
were taken successively at constant field intervals (usually 10 G). Figure 7
shows the induction profiles taken at $T=$ 23 K, while the external field
was ramped down at a rate of 16 G/sec. For external fields between 420 and
240 G the profiles exhibit a break, progressing into the sample interior
with time. In contrast to FSU experiments, here the breaks appear at
approximately the same induction, $B_{f}$ = 360 G, almost independent of the
location in the sample. As before, the breaks reveal coexistence of a
quasi-ordered vortex phase (characterized by low persistent current density)
near the sample edges, and a TDVS (characterized by high persistent current
density), in the sample interior. The value of $B_{f}$ is strongly
suppressed by increasing the sweep rate at constant temperature or
decreasing temperature at a constant sweep rate, as demonstrated in Figures
8 and 9, respectively. Obviously, in these experiments the source of the
transient disordered state cannot be associated with edge contamination, as
flux does not enter the sample through its edges. Moreover, the low $j$
quasi-ordered phase appears near the edges and propagates with time into the
sample interior. The origin of the TDVS is rather 'supercooling' of the
high-field disordered state \cite{vanderBeekSC}. As the field is rapidly
lowered below the transition field, the\ initial thermodynamically
established disordered state is supercooled to inductions below $B_{od}$ and
consequently the apparent solid-solid transition induction, $B_{f}$, shifts
below $B_{od}$. As shown in Figure 8, larger sweep-rates induce 'deeper'
supercooling, shifting $B_{f}$ further down \cite{Kupfer}.

Figure 10 describes the front velocity as a function of time in FSD
experiments at 23 K, for different sweep rates. For low sweep rates, the
front moves at an approximately constant velocity, which depends on the rate
of change of the external field. For high sweep rates, the front movement
accelerates with time.

\section{\protect\bigskip Theoretical analysis}

The experiments described above reveal different dynamic behaviors of the
TDVS depending on the type of experiment (FSU, FSD or constant field), the
rate of change of the external field, the induction at the interface between
the TDVS and the quasi-ordered thermodynamic phase, temperature and time. In
this section we analyze these behaviors in terms of a competition between
two fundamental processes: creation and annealing of TDVS. The parameter
that plays a key role in our analysis is the annealing time, $\tau $, of the
transient disordered vortex state. In defining $\tau $, we refer to the
annealing experiment described in Section IIB. We assume an initial TDVS
throughout the whole sample, created, e.g. by a step increase of the
external field. The annealing process begins\ at $t=0$ with a nucleation of
a quasi-ordered vortex phase at the sample center, and continues with front
propagation of this phase towards the sample edges \cite{GillerPRL}. In
order to distinguish between normal magnetic relaxation and the annealing
process of the TDVS, we assume that the disordered and quasi-ordered vortex
states are characterized by {\em time independent} high and low current
densities, $j_{h\text{ }}$and $j_{l}$, respectively. Such an idealized
annealing process is schematically described in Figure 11a.{\em \ }Let us
examine a location $x$ in the sample where at $t=0$ the induction is $B(x)$
and the current density is $j_{h\text{ }}$. The annealing time, $\tau (B,T)$%
, is defined as the time that it takes for the current density at $x$ to
transform from $j_{h\text{ }}$to $j_{l\text{ \ }}$while the induction at $x$
remains constant and equal to $B$ \cite{note1}. Formally,

\begin{equation}
\tau (B,T)=t_{N}+%
\displaystyle\int %
_{\frac{d}{2}}^{x}\frac{dx}{v_{f\text{ }}},  \label{Definition of tau}
\end{equation}%
where, $t_{N}$ is the nucleation time, $v_{f\text{ }}(B,T)$ is the front
velocity under the conditions of constant external field $H_{ext}$ and
constant $j_{h\text{ }}$and $j_{l}$; $d/2$ denotes the location of the
sample center, where the thermodynamic quasi-ordered vortex phase starts to
nucleate \cite{dimaTheory}. According to Eq.\ref{Definition of tau}, $%
t_{N}=\tau (B_{0}),$ where $B_{0}$ is the induction at the sample center.
Equation \ref{Definition of tau} provides insight into the qualitative
behavior of $\tau (B)$: For low induction $B$, far below the transition
induction $B_{od}$, $v_{f\text{ }}$is large \cite{GillerPRL}, resulting in a
short annealing time $\tau .$ As $B_{od}$ is approached, $v_{f\text{ }}$
decreases causing $\tau $ to increase. In a close vicinity of $B_{od}$, $v_{f%
\text{ }}$approaches zero, and consequently $\tau $ approaches infinity.

\bigskip

In the following subsections we analyze the dynamics of the transient vortex
states in the three experiments described in Section III and show how $\tau $
can be extracted from each of these experiments.

\bigskip

\subsection{Field sweep up}

In these experiments, two competing processes take place simultaneously: A
TDVS is continuously injected into the sample through its edges by the
ramped external field and flux creep, and at the same time the annealing
process takes place. This competition between injection and annealing of the
TDVS determines the position, $x_{f}$, of the front. In the framework of the
critical state model, one can consider $x_{f}$ as a function of three
independent variables \cite{noteREF}: the external field, $H_{ext}$, the
current density, $j_{h}$, and the induction, $B_{f}$, at the front (see
Figure 12a):

\begin{equation}
x_{f}=(H_{ext}-B_{f})/j_{h}.  \label{xfup}
\end{equation}

Thus,

\bigskip

\begin{equation}
\frac{\partial x_{f}}{\partial t}=\frac{1}{j_{h}}\frac{dH_{ext}}{dt}-\frac{%
H_{ext}-B_{f}}{j_{h}^{2}}\left( \frac{\partial j_{h}}{\partial t}\right)
_{H_{ext},B_{f}}-\frac{1}{j_{h}}\left( \frac{\partial B_{f}}{\partial t}%
\right) _{H_{ext},j_{h}}.  \label{vfup}
\end{equation}

According to Eq. \ref{Definition of tau},

\begin{equation}
\left( \frac{\partial B_{f}}{\partial t}\right) _{H_{ext},j_{h}}=\left( 
\frac{\partial B_{f}}{\partial x_{f}}\right) \left( \frac{\partial x_{f}}{%
\partial t}\right) _{H_{ext},j_{h}}=j_{h}v_{f}=1/\left( \partial \tau
/\partial B\right) _{B=B_{f}}.  \label{dbf/dt}
\end{equation}

Thus,

\begin{equation}
\frac{\partial x_{f}}{\partial t}=\frac{1}{j_{h}}\frac{dH_{ext}}{dt}-x_{f}%
\frac{\partial }{\partial t}(\ln j_{h})-\frac{1}{j_{h}}\frac{1}{(\partial
\tau /\partial B)_{B=B_{f}}}.  \label{vfupf}
\end{equation}

The meaning of this equation is as follows: The first term on the right hand
side describes a continuous injection of TDVS by the change of the external
field, pushing the front of the injected TDVS from the sample edge towards
the sample center at a rate determined by the rate of change of $H_{ext}$
(see figure 11b). The second term on the right hand side of Eq. \ref{vfupf}
describes slow injection, in the same direction, caused by flux creep (note
that $\frac{\partial }{\partial t}(\ln j_{h})<0)$ (see figure 11c)$.$The
third term describes the annealing process, pushing the front in the {\em %
opposite} direction, i.e. towards the sample edge (see figure 11a). The
competition between the above three terms determines the dynamics of the
front. Initially, when $H_{ext}$ starts to increase from a zero value, the
inductions involved are far below $B_{od}$, and thus $\partial \tau
/\partial B$ is small (for a schematic description of $\tau (B)$ see figure
13). As a result, the annealing term (third term on the right hand side of
Eq. \ref{vfupf}) dominates, and the TDVS has no chance to propagate into the
sample, i.e. the front is stuck at the sample edge. As $H_{ext}$ increases,
the induction at the edge increases towards $B_{od}$, and the annealing term
becomes\ less and less significant, because the annealing time $\tau $, as
well as $\partial \tau /\partial B$, continuously increase, approaching
infinity as $B\rightarrow B_{od}$. Thus, the front starts its journey from
the sample edge towards its center with a zero velocity, and accelerates
with time towards a final velocity determined by the rate of change of the
external field (neglect of the flux creep term in Eq. \ref{vfupf} is
justified for large enough $dH_{ext}/dt$). We note that $x_{f}$ may reach
the sample center, and thus disappear, before this final velocity is
obtained. Let us denote by $B_{f0}$ the induction at the first detected
front; Since $x_{f0}\approx 0$, substitution of $\partial x_{f}/\partial t=0$
in Eq. \ref{vfupf} yields:

\begin{equation}
(\partial \tau /\partial B)_{B=B_{f0}}=\frac{1}{dH_{ext}/dt}.
\label{dtau/dB}
\end{equation}

Thus, measurements of $B_{f0}$ for different rates of change of the external
field yield $\partial \tau /\partial B$ as a function of $B$. On the basis
of these data one can calculate $\tau $ as a function of $B$ by integration:

\begin{equation}
\tau (B)=%
\displaystyle\int %
_{0}^{B}(\partial \tau /\partial B)dB.  \label{tauintegration}
\end{equation}

Since $\partial \tau /\partial B$ increases monotonically as $B_{od}$ is
approached, Eq. \ref{dtau/dB} implies that $B_{f0}$ is shifted down to lower
inductions as $dH_{ext}/dt$ increases, in accordance with the experiment
(see Figures 2 and 14).

The time dependence of $B_{f}$ can be analyzed by considering $B_{f}$ as a
function of $H_{ext}$, $j_{h}$, and $x_{f}$ (see figure 12a):

\begin{equation}
B_{f}=H_{ext}-j_{h}x_{f}.  \label{Bfup}
\end{equation}

Thus:

\begin{equation}
\frac{\partial B_{f}}{\partial t}=\frac{dH_{ext}}{dt}-x_{f}\frac{\partial
j_{h}}{\partial t}-j_{h}\frac{\partial x_{f}}{\partial t}.  \label{dBf/dt}
\end{equation}

When the front (indicated by a break in the profile) just appears: $%
x_{f}\approx 0$ and $\partial x_{f}/\partial t=0,$ thus $\partial
B_{f}/\partial t=dH_{ext}/dt$, implying that $B_{f}(t)$ starts to rise at a
rate equal to the rate of change of the external field. The second term on
the right hand side of Eq. \ref{dBf/dt}, $-x_{f}(\partial j_{h}/\partial t),$
also contributes to an increase of $\partial B_{f}/\partial t$ with time,
however for large $dH_{ext}/dt$ this contribution is negligible. The
contribution of the third term, $-$ $j_{h}(\partial x_{f}/\partial t),$ is
much more significant and it causes a continuous decrease of $\partial
B_{f}/\partial t$ with time down to zero, which is obtained when $\partial
x_{f}/\partial t$ reaches its final value $(1/j_{h})(dH_{ext}/dt)$ (see Eq. %
\ref{vfupf}). Thus, the increase of $\partial x_{f}/\partial t$ with time is
translated into a decrease of $\partial B_{f}/\partial t$ with time. When
flux injection due to flux creep is negligible, $\partial B_{f}/\partial t=0$
for $B_{f}=B_{od}.$ As noted before regarding $x_{f}$, $B_{f}$ may disappear
before reaching its saturated value $B_{od}$. In this case, the final value
of $B_{f}$ will be $H_{ext}-j_{h}d/2.$

\bigskip

\subsection{Annealing at a constant field}

For a constant $H_{ext}$, Equations \ref{vfupf} and \ref{dBf/dt} become

\begin{equation}
\frac{\partial x_{f}}{\partial t}=-x_{f}\frac{\partial }{\partial t}(\ln
j_{h})-\frac{1}{j_{h}}\frac{1}{(\partial \tau /\partial B)_{B=B_{f}}}
\label{vfaneal}
\end{equation}

and

\begin{equation}
\frac{\partial B_{f}}{\partial t}=-x_{f}\frac{\partial j_{h}}{\partial t}%
-j_{h}\frac{\partial x_{f}}{\partial t}.  \label{Bfaneal}
\end{equation}

In the absence of injection by the change of the external field, injection
by flux creep becomes significant. Thus, the terms $-x_{f}\frac{\partial }{%
\partial t}(\ln j_{h})$ and $-x_{f}\frac{\partial j_{h}}{\partial t}$ in
Eqs. \ref{vfaneal} and \ref{Bfaneal}, respectively, may not be neglected.
Again, two competing processes may be recognized: injection of a TDVS by
flux creep, pushing $x_{f}$ towards the sample center, and annealing of the
TDVS, pushing $x_{f}$ $\ $towards the sample edge. Note that regarding $%
B_{f} $, both processes contribute to an increase of $B_{f}$ with time.
Assuming that initially $x_{f}=d/2$ and $B_{f}$ is far below $B_{od}$, the
second term on the right hand side of Eq.\ref{vfaneal} dominates (i.e. the
annealing process dominates), and the front starts its journey towards the
sample edge with a maximum velocity given by $-\frac{1}{j_{h}}\frac{1}{%
(\partial \tau /\partial B)_{B=B_{f}}}.$ As $x_{f}$ moves towards the sample
edge, $B_{f}$ increases, and consequently $(\partial \tau /\partial
B)_{B=B_{f}}$ increases. Thus, the annealing process slows down, and $x_{f}$
decelerates until $\partial x_{f}/\partial t=0.$ In the absence of the creep
term in Eq. \ref{vfaneal}, $\partial x_{f}/\partial t=0$ would imply $%
B_{f}=B_{od}.$ Obviously, for $B_{f}$ close to $B_{od}$, the injection by
creep becomes significant and the first term on the right hand side of Eq. %
\ref{vfaneal} must be taken into account. This implies that $\partial
x_{f}/\partial t=0$ at an induction $B_{fm}$ {\em smaller} than $B_{od}$
satisfying the equation:

\begin{equation}
-x_{f}\frac{\partial }{\partial t}(\ln j_{h})=\frac{1}{j_{h}}\frac{1}{%
(\partial \tau /\partial B)_{B=B_{fm}}}.  \label{Bfm}
\end{equation}

When this condition is fulfilled, $B_{f}$ continues to increase, beyond $%
B_{fm},$ at a rate $-x_{f}(\partial j_{h}/\partial t)$ due to the flux creep
(see Eq. \ref{Bfaneal}).$\ $For $B_{f}$ larger than $B_{fm},$ the motion of
the front reverses direction, i.e. $\partial x_{f}/\partial t>0$, implying
movement of the front towards the sample center.

\subsection{Field sweep down}

The key difference between field sweep down (FSD) and field sweep up (FSU)
experiments is the source of the TDVS. Unlike FSU experiments, in FSD
experiments the TDVS are not injected through the sample edges, but created
by ''supercooling'' of a previously established thermodynamic disordered
vortex state. We assume that the ramping down of the external field begins
at $t=0$, from a large enough value, $H_{m}$, so that initially the entire
induction profile in the sample is above $B_{od}$. Thus,

\begin{equation}
H_{ext}(t)=H_{m}-\left| \frac{dH_{ext}}{dt}\right| t,  \label{Hdown}
\end{equation}
and the vortices in the entire sample are in a thermodynamic\ disordered
phase from $t=0$ up to $t=t_{od},$

\begin{equation}
t_{od}=\frac{H_{m}-B_{od}}{\left| dH_{ext}/dt\right| },  \label{tTransient}
\end{equation}
when the induction at the sample edge equals $B_{od}$. For $t>t_{od}$, a
supercooled TDVS begins to appear at the sample edge; as the external field
continues to drop, the front of the supercooled transient states penetrates
deeper into the sample. At the same time the annealing process takes place.
Note that in contrast to the annealing experiments (Section IIIB), in FSD
experiments the annealing process starts at the sample {\em edge,\ }where
the induction is smallest, and thus the lifetime of the supercooled TDVS is
shortest. Clearly, it will require infinitely long time to anneal the TDVS
generated at the edge immediately after $t=t_{od}$. However, as the external
field is ramped down, the induction at the edge drops, and consequently the
annealing time continuously decreases. Thus, at a later time $t_{f0}>t_{od},$
the induction at the edge drops to $B_{f0}=$ $B_{od}-\left|
dH_{ext}/dt\right| (t_{f0}-t_{od})$, and consequently, the annealing time
decreases to $\tau (B_{f0})$. A quasi-ordered vortex state (and thus the
first front) will appear at the sample edge when

\begin{equation}
\tau \left( B_{f0}\right) =t_{f0}-t_{od}=\frac{B_{od}-B_{f0}}{\left|
dH_{ext}/dt\right| }.  \label{taudown}
\end{equation}

This condition defines $B_{f0}$ at the crossing point of\ the\ straight line 
$(B_{od}-B)/\left| dH_{ext}/dt\right| $ plotted versus $B,$ and the curve $%
\tau (B)$, (see Figure 13). From the shape of the $\tau (B)$-curve it is
clear that as $\left| dH_{ext}/dt\right| $ increases, this crossing point is
shifted towards lower inductions $B$. This is in accordance with the
experiment, which shows that as the ramping rate increases the induction at
the first detected break in the induction profile is shifted down (see
Figures 8 and 14). On the basis of Eq. \ref{taudown}, one can generate the $%
\tau $ vs. $B$ curve by measuring $B_{f0}$ for different ramping rates of
the external field. We note, however, that in this case a pre-determination
of $B_{od}$ is required.

Once the front appears at the sample edge, its dynamics (i.e. $x_{f}$ and $%
B_{f}$ versus time) can be analyzed starting from an equation similar to Eq. %
\ref{xfup} (see Figure 12b):

\begin{equation}
x_{f}=\frac{B_{f}-H_{ext}}{j_{l}}.  \label{xfdown}
\end{equation}

Using the same arguments as in the analysis of the FSU experiments, one
obtains:

\begin{equation}
\frac{\partial x_{f}}{\partial t}=-\frac{1}{j_{l}}\frac{dH_{ext}}{dt}-x_{f}%
\frac{\partial }{\partial t}(\ln j_{l})+\frac{1}{j_{l}}\frac{1}{(\partial
\tau /\partial B)_{B=B_{f}}}  \label{vfdown}
\end{equation}

and

\begin{equation}
\frac{\partial B_{f}}{\partial t}=\frac{dH_{ext}}{dt}+x_{f}\frac{\partial
j_{l}}{\partial t}+j_{l}\frac{\partial x_{f}}{\partial t}.
\label{dBf/dtdown}
\end{equation}

It is interesting to note that as far as $x_{f}$ is concerned, in FSD
experiments the three factors involved, namely ramping down of $H_{ext}$,
flux creep, and the annealing, all push $x_{f}$ in the same direction -
towards the sample center. In contrast, the dynamics of $B_{f}$ is governed
by competing processes: both, ramping down of $H_{ext}$ and flux creep push $%
B_{f}$ downwards ($dH_{ext}/dt<0$, and $\partial j_{l}/\partial t<0$), while
the annealing process shifts $B_{f}$ upwards ($\partial x_{f}/\partial t>0$%
). Thus, variations with time of $x_{f}$ relative to $B_{f}$ are much larger
in FSD as compared to FSU experiments. This explains the outstanding
qualitative difference between the induction profiles measured in FSD and
FSU experiments, namely the apparent constancy of $B_{f}$ in FSD experiments
as compared to FSU experiments.

In order to follow the behavior of $x_{f}$ and $B_{f}$ with time, let us
first ignore the contribution of flux creep (second term on the right hand
sides of Eqs. \ref{vfdown} and\ \ref{dBf/dtdown}). For relatively slow rate
of change of $H_{ext},$ the first front appears at a relatively high
induction, thus $(\partial \tau /\partial B)_{B=B_{f0}}$ is large, and the
third term on the right hand side of Eq. \ref{vfdown} may be neglected
compared to the first term. In this case $\partial x_{f}/\partial t\approx
(-1/j_{l})(dH_{ext}/dt).$ Thus, the initial velocity of the front is
determined by the rate of change of the external field, in contrast to FSU
experiment where the {\em final} velocity of the front is determined by $%
dH_{ext}/dt$. Furthermore, Eq. \ref{dBf/dtdown} implies that in this case $%
\partial B_{f}/\partial t\approx 0$, i.e., $B_{f}$ is constant. The negative
creep term in Eq. \ref{dBf/dtdown} causes slow decrease of $B_{f}$ with
time. According to Eq. \ref{vfdown}, the slow decrease of $B_{f}$ with time
causes slow increase of $\partial x_{f}/\partial t$ with time. For high rate
of change of $H_{ext},$ $B_{f0}$ is small, and consequently $1/(\partial
\tau /\partial B)_{B=B_{f0}}$ may not be neglected compared to $dH_{ext}/dt$%
. In this case, the term $(1/j_{l})(1/(\partial \tau /\partial B)_{B=B_{f0}}$
contributes to increase the velocity of $x_{f}$. Similar contribution in the
same direction is obtained from the creep term.

\section{Discussion}

\bigskip The above analysis shows that the dynamics of TDVS is determined by
the rate of change of the external field, the creep rate, and the rate of
change with $B$ of the annealing time $\tau (B)$. While the external field
is changed at a constant rate, and the creep rate is slow, the annealing
rate varies over a wide range, due to the strong dependence of $\tau $ on $%
B. $ This dependence can be extracted from FSU and FSD experiments, using
Eqs. \ref{dtau/dB} and \ref{taudown}, respectively, and can be measured
directly from annealing experiments in the range of $B$ where injection by
creep can be neglected, as discussed in details in ref \cite{beenaBrief} .
For the extraction of $\tau $ from FSU experiments, one only needs to
measure $B_{f0} $ for different sweeping rates of $H_{ext}.$ The circles in
Figure 14 show results of measurements of $B_{f0}$ as a function of $%
dH_{ext}/dt$ in FSU experiments conducted at $T=23$ K. As predicted
theoretically (see Section IIIA), $B_{f0}$ decreases monotonically as $%
dH_{ext}/dt$ increases. Results for $\tau (B)$ extracted from these data,
using Eq. \ref{dtau/dB}, are shown by the circles in Figure 15. The solid
line in this figure shows a theoretical fit to the Eq.

\begin{equation}
\tau =\frac{\tau _{0}}{(1-\frac{B}{B_{od}})^{\gamma }}.  \label{tau}
\end{equation}%
Evidently, a good fit is obtained over a wide range down to $B=200$ $G.$ The
fit yields $\tau _{0}=0.011$ $s$, $\gamma =2.6$ and $B_{od}=460$ $G$. This
value of $B_{od}$ is used for the extraction of $\tau (B)$ from FSD
experiments. The squares in Figure 14 show $B_{f0}$ as a function of $%
dH_{ext}/dt$ in FSD experiments at the same temperature. Interestingly, one
observes that the values of $B_{f0}$ in FSU (circles) and FSD (squares)
experiments are similar. Note that for $\gamma =1$ in Eq. \ref{tau}, Eqs. %
\ref{dtau/dB} and \ref{taudown} predict the same\ values of $B_{f0}$ for
both experiments for a given $dH_{ext}/dt$. For $\gamma >1$, these equations
predict higher values of $B_{f0}$ for FSD experiments \cite{note2}, as
observed experimentally. The bold squares in Fig. 15 show $\tau (B)$ as
extracted from FSD experiments\ using Eq. \ref{taudown}. Evidently, these
results are in a very good agreement with the results of measurements of $%
\tau (B)$ from FSU experiments (circles in Fig. 15). The triangles in Fig.
15 depict direct measurements of $\tau (B)$ from annealing experiments \cite%
{beenaBrief}. Good agreement with the previous results is obtained over a
wide range of inductions. Deviations are expected for larger inductions ,\
where injection of TDVS by creep becomes significant.

Measurements of $\tau (B)$ at different temperatures are shown in Figure 16.
Evidently, the lifetime spectrum of the TDVS widens as temperature is
lowered \cite{beenaRapid}. Fits of these data to Eq. \ref{tau}, yields $\tau
_{0}(T)$ and the thermodynamic transition line $B_{od}(T)$. In Figure 17 we
show the temperature dependence of $B_{od}$ as extracted by this method. For
comparison, we also show non-equilibrium transition lines \cite{beenaRapid}
(dashed curves) as measured by the onset of the second magnetization peak,
for different sweeping rates of the external field. It is seen that
deviations from the thermodynamic transition line increase with increasing
sweeping rates. Direct measurement of the thermodynamic transition line
would require measurements at extremely low rates to allow for complete
annealing of the TDVS. Alternatively, one can determine $B_{od}(T)$
indirectly from Eq. \ref{tau}, by measuring $\tau (B,T)$ as outlined above.

Figure 18 shows the temperature dependence of $\tau _{0}$ as obtained from
the theoretical fits of the data of Fig. 16 to Eq. \ref{tau}. Apparently, $%
\tau _{0}$ increases exponentially as the temperature is lowered. The solid
line in this figure is a fit of $\tau _{0}(T)$ to Arrhenius law: $8\times
10^{-9}\exp (326/T)$. The exponential increase of $\tau _{0}$ as temperature
is lowered explains the disappearance of the second magnetization peak at
low temperature, which was misinterpreted as termination of the transition
line \cite{beenaRapid,beenaBrief,Yesh-Bont,Wen}.

Once the functional dependence of $\tau $ on $B$ is known, one can test the
predictions of the above analysis for the velocity $\partial x_{f}/\partial
t $ of the interface between the quasi-ordered phase and the TDVS, in the
three experiments described above. The solid curves in Figures 4 and 10 are
calculated from Eqs. \ref{vfupf} and \ref{vfdown} using the measured values
of $j_{h}(t),$ $j_{l}(t)$ and $B_{f0}$. A fairly good agreement with the
experimental data is obtained using $B_{od}=480$ $G$.

A totally different behavior of $\partial x_{f}/\partial t$\ is obtained in
annealing experiments. As described in Section IIB, in this case the initial
velocity is high, it gradually drops to zero at an induction below $B_{od}$,
and then reverses direction (see Figure 6). The analysis outlined above (see
Section IIIB) is capable of predicting this behavior as shown by the solid
curve in Figure 6. This curve was calculated taking into account both, the
annealing term and the creep term in Eq. \ref{vfaneal}. A reasonable
agreement with the experimental data is obtained using the same value of $%
B_{od}$, and $\tau _{0}=0.04$ $s$.

\section{Summary and conclusions}

We have visually traced the processes of generation and annealing of TDVS,
using a high-speed magneto-optical system. Snapshots of the induction
distribution reveal front-like propagation or retreat of the transient
disordered states. The velocity of this front is governed by the rate of
change of the external field, the induction at the front, temperature and
time. We analyzed the dynamics of the front, in three different experiments,
in terms of competition between creation and annealing of TDVS, where the
annealing process is governed the annealing time $\tau (B,T)$. The
predictions of this analysis are in good agreement with the experimental
results. In particular, extraction of the annealing time $\tau (B,T)$ from
the three experiments, yield the same results. The exponential increase of $%
\tau (B)$ with decreasing temperature, explains broadening of the second
magnetization peak and its disappearance below a certain temperature \cite%
{beenaRapid,beenaBrief,Yesh-Bont,Wen} - a phenomenon that was misinterpreted
as termination of the transition line. Reliable measurement of the
thermodynamic order-disorder transition line $B_{od}(T)$ requires magnetic
measurements at extremely low field sweep rates to allow for complete
annealing of the transient states. This requirement becomes more severe as
the temperature is lowered, because of the exponential increase of the
annealing time. The thermodynamic transition line, $B_{od}(T)$, can be
indirectly determined by fitting $\tau $ to Eq. \ref{tau}. Measurement of
the transition line in this method yields results which are significantly
different from the non-equilibrium transition lines, commonly measured from
the onset of the second magnetization peak, neglecting effects of transient
vortex states.

\bigskip

\section{\protect\bigskip Acknowledgements}

We thank Tsuyoshi Tamegai for providing us with the Bi$_{2}$Sr$_{2}$CaCu$%
_{2} $O$_{8+\delta }$ crystal. A.S. acknowledges support from the
German-Israel Foundation (GIF). Y.Y. acknowledges support from the ISF
Center of Excellence Program, by the Heinrich Hertz Minerva Center for High
Temperature Superconductivity, and by the Wolfson Foundation.

Figure 1. Induction profiles across a BSCCO crystal measured at $T=$ 23 K,
while ramping the external field at a rate of 53 G/sec. Note the sharp
change in the slope of the profiles (''break'') appearing for $H_{ext}>430$
G. Bold lines represent profiles with and without (shorter times) break.

Figure 2. Induction profiles measured\ at $T=$ $25$ $K$ while ramping the
external field at different rates. As $dH_{ext}/dt$ increases, the first
break in the profiles appears at a lower induction.

Figure 3. Induction profiles measured\ at different temperatures, while
ramping the external field at the same rate. As temperature is lowered, the
first break in the profiles appears at a lower induction.

Figure 4. Velocity of the penetrating front of the transient disordered
state as a function of time for different sweep rates of the external field.
Solid lines are calculated from Eq. \ref{vfup}.

Figure 5. Time evolution of the induction profiles after abruptly increasing
the external field from zero to 465 G. At approximately t = 0.5 s, a sharp
change (a 'break') in the slope of the profiles appears, progressing with
time towards the sample edge.

Figure 6. Velocity of the propagating front of the quasi-ordered vortex
phase as a function of time for $H_{ext}$ $=$ $500$ and $625$ $G$. Note that
the movement of the front reverses its direction, indicating an end of the
annealing process and beginning of injection of transient disorder vortex
state by flux creep. Solid lines are calculated from Eq. \ref{vfaneal}.

Figure 7. Induction profiles measured at $T=$ $23$ $K$, while the external
field was ramped down at a rate of 16 G/sec. For external fields between 420
and 240 G the profiles exhibit a break, progressing into the sample interior
with time. Bold lines represent profiles with and without (longer times)
break.

Figure 8. Induction profiles measured\ at $T=$ $25$ $K$ while ramping down
the external field at different rates. As $\left| dH_{ext}/dt\right| $
increases, the breaks in the profiles appear at lower inductions.

Figure 9. Induction profiles measured\ at $T=$ $25$ $K$ and $T=$ $30$ $K$
while ramping down the external field at 160 G/sec. As temperature
decreases, the breaks in the profiles appear at lower inductions.

Figure 10. Velocity of the quasi-ordered phase front as a function of time
in field sweep down experiments, for different sweep rates. Solid curves are
calculated from Eq. \ref{vfdown}.

Figure 11. Schematic induction profiles illustrating motion of the interface
between the quasi-ordered phase and the transient disordered state due to
changes with time of the induction $B_{f}$ $\ $(a), the external field (b),
and the current density $j_{h}$ (c). Diagram (a) illustrates an 'ideal'
annealing process.

Figure 12. Schematic induction profile illustrating the relationship between
the external field $H_{ext}$, the induction $B_{f}$ $\ $at the break, the
current density $j_{h}$ of the transient disordered state, and the location $%
x_{f}$ of the interface between the quasi-ordered phase and the transient
disordered state, for field-sweep-up (a) and field-sweep-down (b).

Figure 13. Graphic procedure for determination of the induction $B_{f0}$ at
the first detected interface between the quasi-ordered phase and the
transient disordered state, in field sweep down experiments: $B_{f0}$ is the
crossing point of\ the\ straight line $(B_{od}-B)/(dH_{ext}/dt)$ plotted
versus $B,$ and the curve $\tau (B).$

Figure 14. Measurements of $B_{f0}$ as a function of $dH_{ext}/dt$ in field
sweep up (circles) and field sweep down (squares) experiments conducted at $%
T=23$ $K$.

Figure 15. Measurements of the annealing time $\tau $ as a function of $B$
from field sweep up (circles), field sweep down (squares) and annealing
(triangles) experiments. The solid line is a theoretical fit.

Figure 16. Measurements of the annealing time $\tau (B)$ at different
temperatures.

Figure 17. The thermodynamic transition line $B_{od}(T)$ (continuous curve)
as obtained from theoretical fit of the data of Fig. 15 to Eq. \ref{tau}.
For comparison, non-equilibrium transition lines, measured from the onset of
the second magnetization peak for different field sweep rates, are shown by
the dashed lines.

Figure 18. $\tau _{0}(T)$ as obtained from theoretical fit of the data of
Fig. 15 to Eq. \ref{tau}.


\begin{references}
\bibitem{note1} Our definition of $\tau $ may imply dependence of $\tau $ on
the location $x$. However, measurements of $\tau $ at different locations in
BSCCO samples \cite{beenaRapid} show only minor variations in $\tau $ at
different locations with the same $B.$
\bibitem{noteREF} Our analysis strictly applies to the parallel geometry
where demagnetization effects are absent. Including these effects for the
perpendicular geometry is not expected to change the overall physical
picture of competition between injection and annealing: Obviously, the
annealing process will not be affected, as it is governed by $\tau (B)$; For
the injection process, the role of $dH_{ext}/dt$ will be played by $%
dB_{edge}/dt$.
\end{references}
\end{document}